\documentclass{PoS}

\newcommand{\F}{\textit{Fermi}}
\newcommand{\g}{$\gamma$}
\newcommand{\hi}{\mathrm{H\,\scriptstyle{I}}}
\newcommand{\zm}{$z_\mathrm{max}$}
\newcommand{\apj}{Astroph. J.}
\newcommand{\apjs}{Astroph. J. Suppl. S.}
\newcommand{\aap}{Astron. Astroph.}
\newcommand{\mnras}{Mon. Not. R. Astron. Soc.}

\title{Tracing the propagation of cosmic rays in the Milky Way halo with \F-LAT observations of high- and intermediate-velocity clouds}

\ShortTitle{\F-LAT observations of HVCs and IVCs}

\author{\speaker{L. Tibaldo} and S.~W. Digel on behalf of the \F-LAT collaboration\\
        Kavli Institute for Particle Astrophysics and Cosmology, SLAC National Accelerator Laboratory, Menlo Park, CA 94025, USA\\
        E-mail: \email{ltibaldo@slac.stanford.edu},  \email{digel@stanford.edu}}

\abstract{Cosmic rays up to at least PeV energies are usually described in the framework of an elementary scenario that involves acceleration by objects that are located in the disk of the Milky Way, such as supernova remnants or massive star-forming regions, and then diffusive propagation throughout the Galaxy. Details of the propagation process are so far inferred mainly from the composition of cosmic rays measured near the Earth and then extrapolated to the whole Galaxy. The details of the propagation in the Galactic halo and the escape into the intergalactic medium remain uncertain. The densities of cosmic rays in specific locations can be traced via the \g~rays they produce in inelastic collisions with clouds of interstellar gas. Therefore, we analyze 73 months of \F-LAT data from 300~MeV to 10~GeV in the direction of several high- and intermediate-velocity clouds that are located in the halo of the Milky Way. These clouds are supposed to be free of internal sources of cosmic rays and hence any \g-ray emission from them samples the large-scale distribution of Galactic cosmic rays. We evaluate for the first time the \g-ray emissivity per hydrogen atom up to ${\sim} 7$~kpc above the Galactic disk. The emissivity is found to decrease with distance from the disk, which provides direct evidence that cosmic rays at the relevant energies originate therein. Furthermore, the emissivity of one of the targets, the upper intermediate-velocity Arch, hints at a 50\% decline of the cosmic-ray intensity within 2~kpc from the disk.}

\FullConference{The 34th International Cosmic Ray Conference,\\
		30 July- 6 August, 2015\\
		The Hague, The Netherlands}

\begin{document}

\section{Introduction}
This conference paper summarizes the main contents of~\cite{HVCpaper}, which presents the first direct constraints on the densities of cosmic-ray (CR) nuclei in the halo of the Milky Way at large scales, from observations in the \g-ray band of high-velocity clouds (HVCs) and intermediate-velocity clouds (IVCs) located at up to 7~kpc from the Galactic plane. 

Several lines of evidence indicate that CRs up to at least PeV energies are produced by energetic objects in the disk of the Milky Way, and after escaping from the accelerators they remain confined by magnetic fields in its halo. The Galactic origin is supported both by multiwavelength observations of supernova remnants and massive-star forming regions that pinpoint the presence of accelerated particles, and by observations of external galaxies in radio and \g~rays which reveal diverse CR environments related to the star-formation properties of each individual galaxy.

Historically, details of CR propagation in the Milky Way have been inferred mainly from elemental and isotopic abundances of CRs directly detected near the Earth. Stable and unstable isotopes rare in nature that are produced by CR interactions in the interstellar medium (ISM) encode the history of the propagation in the Galaxy. Extrapolating the propagation properties inferred from local CRs to the whole Galaxy leads to models that can reproduce with reasonable accuracy all the observables related to CRs, e.g., \cite{strong2007}. 

However, the details of the propagation in remote regions of the Galaxy, including the halo, are uncertain. The vertical extent of the halo has been the subject of notable debate over the past years. Isotopic and elemental abundances give typical constraints on the ``point of no return'', beyond which CRs leave the Galaxy, of $4-6$~kpc from the disk, e.g., \cite{strong2007}. Recently, an analysis of large-scale synchrotron emission from CR electrons in the Milky Way indicated larger heights of ${\sim} 10$~kpc \cite{orlando2013}.

In addition to synchrotron emission, the remote detection of CR nuclei is made possible by \g~rays produced in inelastic collisions with interstellar gas. Therefore, \g-ray observations by the \F\ Large Area Telescope (LAT) have been used in a wealth of studies relevant to understanding CR propagation, e.g., \cite{LATdiffpapII}. Interestingly, a halo height of ${\sim} 10$~kpc offered a possible solution to the \emph{CR gradient problem} revealed by LAT observations, i.e., CR densities in the region of the Milky Way beyond the solar circle greater than expected based on the distribution of putative CR sources \cite{abdo2010cascep,ackermann20113quad}.

In the past, studies of CR nuclei through \g-ray observations were limited to the Galactic disk, where most of the gas resides. However, the \F-LAT survey has enabled us to search for the first time with sufficient sensitivity for \g-ray emission produced by tenuous clouds of gas in the halo of the Milky Way. These are known as HVCs and IVCs because the Doppler-shift velocity of the $\hi$ 21-cm line is inconsistent with Galactic rotation, with HVCs conventionally defined as having velocities with absolute value $>80$--100~km~s$^{-1}$. We selected a subsample of HVCs and IVCs that are massive enough to be conceivably detected by the LAT and that have well defined upper and lower limits on their distance (\emph{distance bracket}) from spectroscopic measures of foreground and background stars \cite{wakker2001}.  

\section{Data analysis}

The analysis method we used relies on the assumption that CR densities have mild variations on the scales of the interstellar cloud complexes that we aim to study, which are located away from potential accelerators. Therefore, the \g-ray intensities from CR-gas interactions can be approximated as the product of gas column densities in each complex and phase of interstellar gas with a \g-ray emissivity per H~atom, i.e., the \g-ray emission rate per H~atom, which is a tracer of CR densities.

In order to determine the \g-ray emissivities, and therefore constrain the CR densities in target HVCs and IVCs, we need to determine the column densities of interstellar gas separately for the complexes located along each line of sight. The selected target HVCs and IVCs are grouped in three independent regions for the analysis.

\subsection{Preparation of interstellar gas maps}

HVCs and IVCs are best traced by the 21-cm line of atomic hydrogen, $\hi$. The Doppler shifts of the lines enables the separation of local gas from HVCs and IVCs along the same line of sight. We used data from the LAB survey \cite{kalberla2005} to determine the $\hi$ column densities in velocity ranges corresponding to the different complexes in each region. In order to most accurately determine the column densities in each of the complexes, the boundaries of the velocity ranges were adjusted to fall at a minimum or inflection point of the $\hi$ temperature profiles, and we performed fits of the profiles using Gaussian functions to correct for cross contamination between adjacent ranges.

The local interstellar medium (ISM) is also rich in molecular gas, which can be traced by the 2.6-mm line of CO. We made use of the CfA CO survey \cite{dame2001} supplemented by observations at high Galactic latitude \cite{devries1987}.  

Finally, a linear combination of  $\hi$ and CO maps does not properly account for the totality of neutral gas. There is an additional component, the dark neutral medium (DNM), which at large scales is best traced by interstellar \g-ray emission and dust thermal emission or extinction \cite{grenier2005}. We built maps of the DNM starting from the maps of dust radiance (or optical depth) derived from \textit{IRAS} and \textit{Planck} observations \cite{planck2013dustmaps} and subtracting the best-fit contribution from the $\hi$ and CO maps. We used an iterative procedure to acknowledge and address the missing DNM component in the fit of the dust maps and therefore obtain an unbiased estimate of the DNM column densities.

\begin{figure}
\begin{center}
  \includegraphics[width=1.\textwidth]{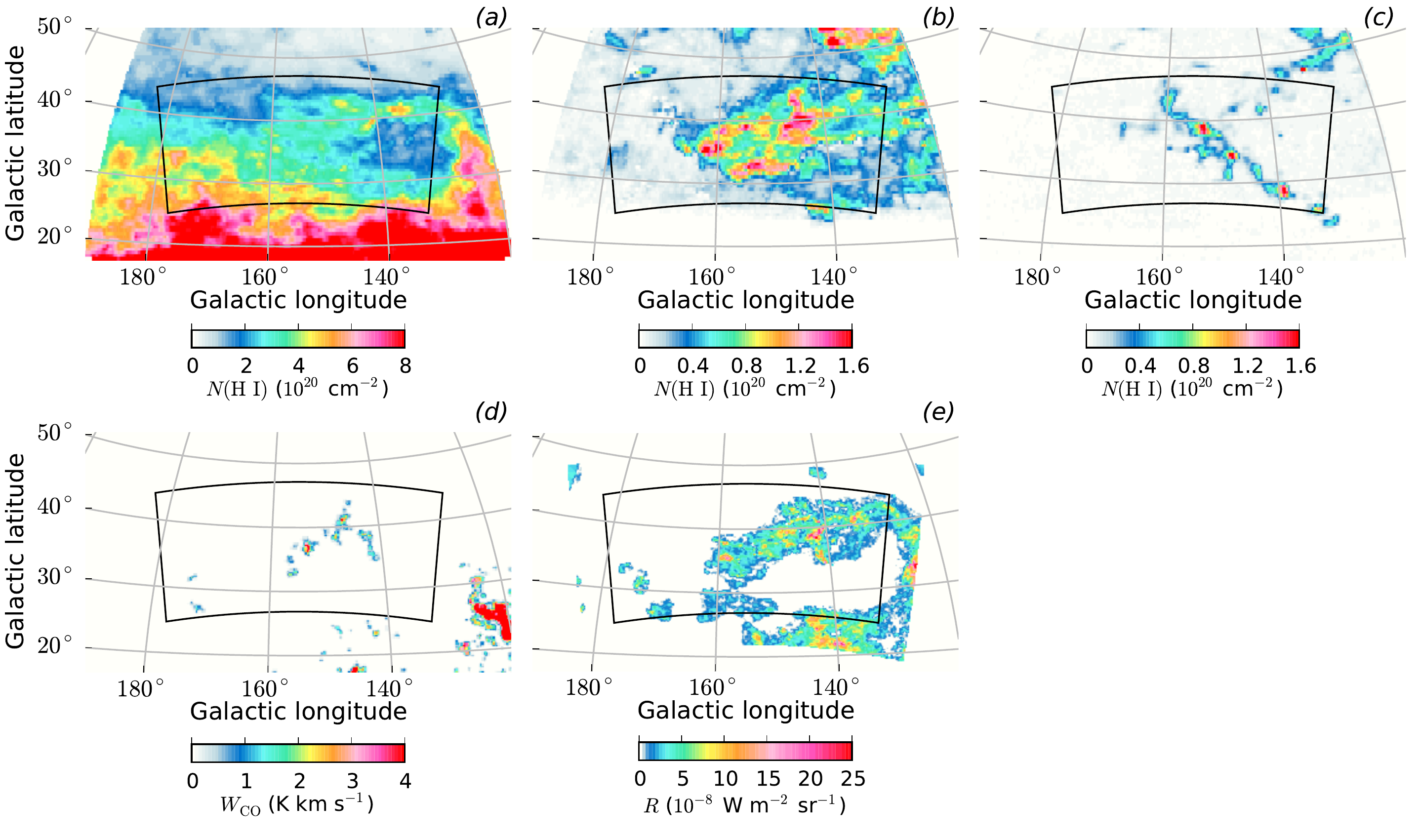}
     \caption{Example of the maps of interstellar gas for one of the regions studied. Maps a, b, and c are column densities of $\hi$ at a) low velocities (local ISM), and for two targets at b) intermediate velocities (the low-latitude intermediate velocity Arch) and c) high velocities (Complex A). The other maps trace additional gas components in the local ISM within a few hundred pc from the solar system. Map d gives the CO line intensity which is a tracer of molecular gas column densities in the Ursa Major complex. Map e is dust radiance detected in excess of the contributions represented by $\hi$ and CO, tracing the DNM in the North Celestial Pole Loop.}
     \label{fig1}
\end{center}
\end{figure}

Fig.~\ref{fig1} shows as an example the maps of interstellar gas derived for one of the regions studied, in the direction of the low-latitude intermediate velocity Arch and of the HVC known as Complex A. The maps illustrate how HVCs and IVCs have distinct morphologies which allows us to separate the \g-ray emissions of the foregrounds and derive the CR densities in those objects.

\subsection{Gamma-ray analysis} 

The \g-ray analysis employs the maps of interstellar gas just described, and assumes that \g-ray intensities are proportional to them. Furthermore, we make the assumption that the spectrum of the \g-ray emission is the same as the one in the local ISM \cite{casandjian2015} and we allow for each map only a global scaling factor. For $\hi$ maps the scaling factor is in fact the ratio of the emissivity in each complex over the emissivity in the local ISM, hence a measure of the ratio of average CR densities in the relevant energy range. For CO and excess dust maps the scaling factor also incorporates the ratio of CO brightness and dust radiance (or optical depth) over column density of H~atoms. We also employ models of additional smoother diffuse components, inverse-Compton emission and the isotropic \g-ray background, as well as of point sources \cite{3FGL}. All diffuse sources and bright point sources have free normalizations, and the inverse-Compton model also has a log-parabolic spectral correction.

The model so defined was fit separately for the three region to 73 months of \texttt{P7REP\_CLEAN} \F~LAT data. We performed a binned likelihood analysis with  Poisson statistics in the energy range from 300~MeV to 10~GeV. The energy range was chosen because the \g-ray emission from gas is largely dominated by nuclear interactions, it provides a good angular resolution and large counting statistics, it is largely free from contamination by \g-ray emission from the Earth's atmosphere, and limits the impact from the choice of the spectral models for the different components adopted for the analysis.

For target HVCs and IVCs we assessed whether there is significant \g-ray emission using the likelihood ratio test. For detected targets (and local complexes as well) we determined the emission spectrum over 6 separate energy bins and verified that it is consistent with the initial assumption \cite{casandjian2015} within statistical uncertainties. For undetected targets we determined a 95\% confidence level (c.l.) upper limit on the emissivity based on a semi-Bayesian integration of the likelihood profile. 

Furthermore, we gauged systematic uncertainties on the results using two independent and complementary approaches. On one hand, we determined the impacts of some modeling choices by varying some of the most important and uncertain inputs to the analysis model: the DNM maps (considering alternatively dust radiance and optical depth as dust tracers, and making different assumptions about the errors in their fits), the models of inverse-Compton emission (varying some parameters relevant to their calculation), and the spectrum of undetected target interstellar clouds (based on an educated guess from CR propagation models).  In addition, we performed jackknife tests repeating the analysis 150 times for each region, each time masking 20\% of the region, in order to investigate the impact of the assumption that CR densities are uniform within each complex and ISM phase and and to investigate whether any particular sub-regions drive the fits  (e.g., due to mismodeled point sources). 

\begin{figure}
\begin{center}
  \includegraphics[width=1.\textwidth]{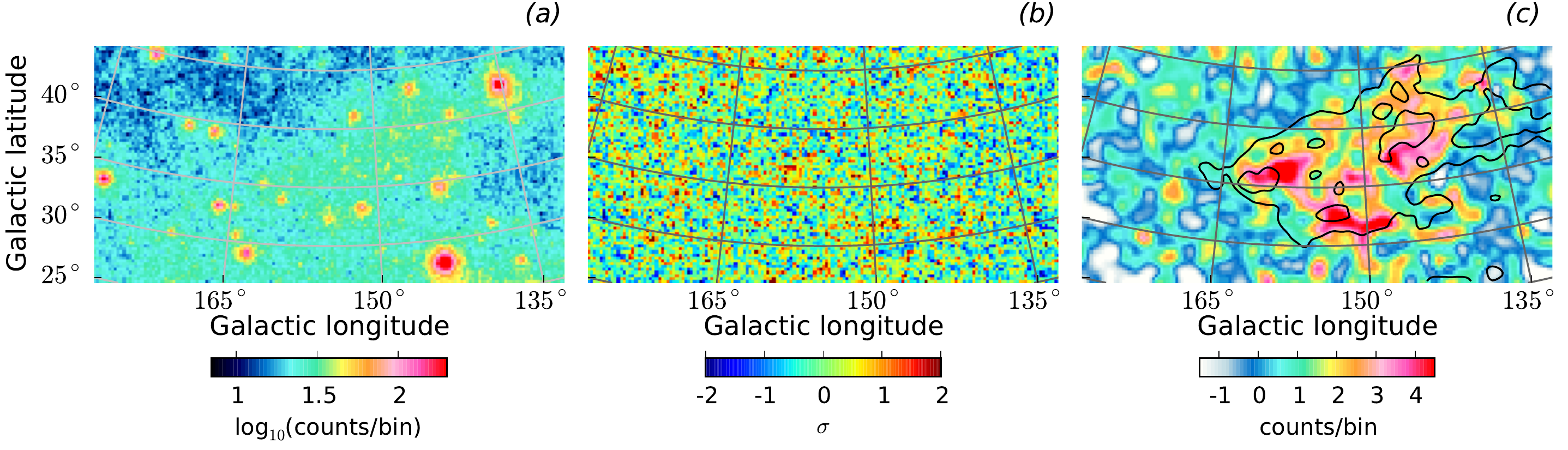}
     \caption{Demonstration of the analysis procedure for one of the regions studied. a) Total number of \g~rays over the energy range from 300~MeV to 10~GeV. b) Residuals after subtraction of the best-fit model. c) Residuals after subtraction of the best-fit model except for the component associated with the IVC known as the low-latitude intermediate velocity arch. The contours show the column densities of interstellar hydrogen in this IVC at $0.6 \times 10^{20}$~atoms~cm$^{-2}$ and $1.2 \times 10^{20}$~atoms~cm$^{-2}$.}
     \label{fig2}
\end{center}
\end{figure}

Further details on the analysis are available in~\cite{HVCpaper}. Fig.~\ref{fig2} illustrates the analysis procedure and shows for an example region how the analysis procedure satisfactorily reproduces the LAT data. In the following we will briefly discuss the main results and their implications.

\section{Results and discussion}

\begin{figure}
\begin{center}
  \includegraphics[width=1.\textwidth]{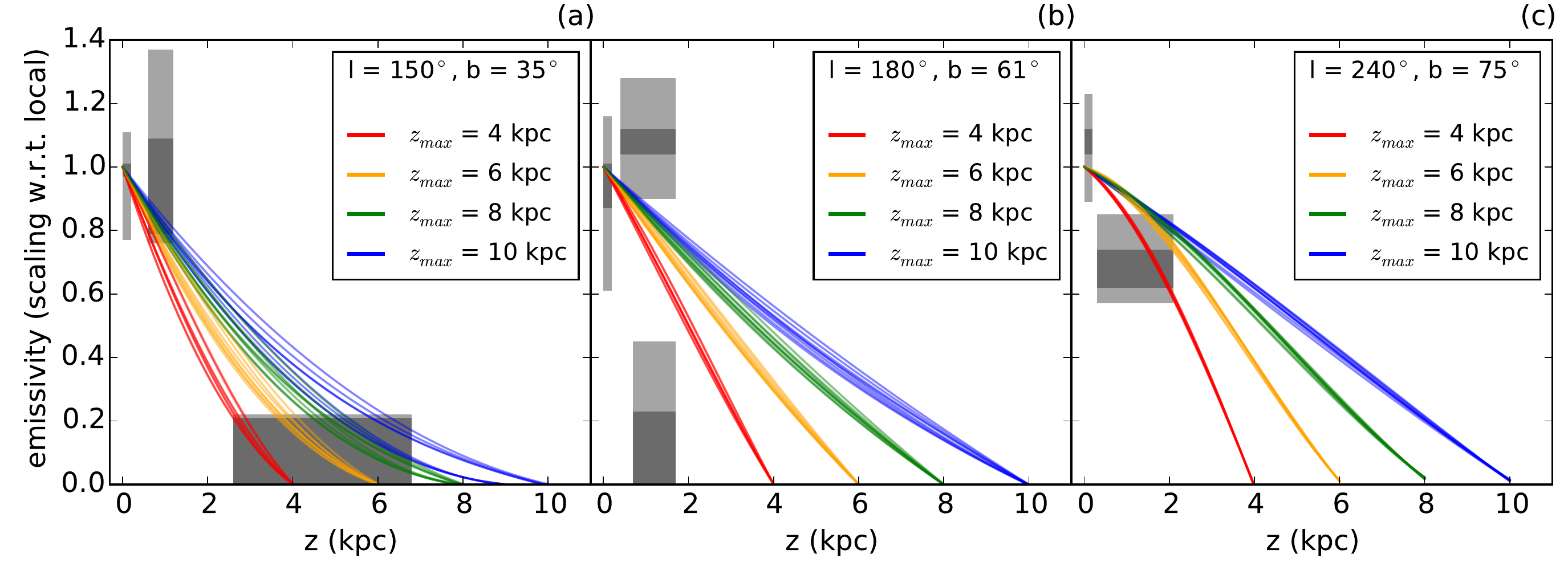}
     \caption{We compare for the three regions studied the emissivity 
scaling factors obtained from LAT data in the 300~MeV to 10~GeV energy range (gray rectangles) with predictions 
from the models in \cite{LATdiffpapII} (curves). 
The horizontal widths of the rectangles indicate the $z$ brackets of 
target IVCs and HVCs, i.e., the range between lower and upper limits on their 
altitudes \cite{wakker2001}. The 
dark gray rectangles have vertical size corresponding to the statistical 
uncertainties, while for the light gray rectangles the vertical size 
encompasses the
combination of statistical and systematic uncertainties. The emissivity of local gas 
is assigned to the range from $z=0$~kpc to $z=0.3$~kpc (disk). The model curves 
from 
\cite{LATdiffpapII} were calculated for the line of 
sight indicated in the legend of each panel, approximately 
corresponding to the column density peaks of the target 
complexes. The curves are color-coded based on the 
maximum heights \zm\ of the CR confinement halo in the 
models.}
     \label{fig3}
\end{center}
\end{figure}

We achieved the first significant detections of \g-ray emission from clouds with known locations in the halo of the Milky Way: the low-latitude intermediate-velocity Arch, the lower intermediate-velocity Arch, and the intermediate-velocity Spur. Fig.~\ref{fig3} summarizes the results, including upper limits for the emissivities of undetected targets.  

Fig.~\ref{fig3} suggests a general trend of decreasing emissivity with increasing distance from the Galactic plane $z$. In region~(b) two clouds have similar altitude brackets, $0.4-1.7$~kpc and $0.7-1.7$~kpc, but different emissivities, consistent with the local value in the first case and $<50\%$ with respect to the local value in the second case. This is not necessarily contradictory, since brackets are pairs of lower and upper limits on the cloud distance, and hence the two objects may have altitudes that differ by more than 1~kpc.

The \g-ray emissivities in the 300~MeV to 10~GeV energy range are a tracer of CR densities in the energy range from ${\sim} 3$~GeV/nucleon to ${\sim} 200$~GeV/nucleon. We evaluated through the Kendall correlation test that there is evidence at 97.5\% c.l. that the emissivities, hence the CR densities, decrease with increasing distance from the plane. This corroborates the notion that CRs in this energy range are accelerated in the disk of the Milky Way and then propagate in its halo, for the first time from directly tracing the CR densities in the halo itself.

In Fig.~\ref{fig3} we also compare the results to predictions from a set of CR propagation models based on the GALPROP code \cite{LATdiffpapII}. The GALPROP input parameter with the largest impact on the vertical gradient of CR densities is the maximum height of the confinement halo \zm. There is broad agreement between emissivities derived from the LAT data and model predictions. In the context of the models considered, the upper limit for the emissivity of the  upper intermediate-velocity Arch being 50\% of the local value is pointing toward a \zm\ value smaller than some values proposed in recent years ranging up to 10~kpc. The low measured emissivity seems to disfavor a large \zm\ as a possible explanation for the \emph{CR gradient problem} in the outer Galaxy \cite{ackermann20113quad}.

We note, on the other hand, that the models considered in Fig.~\ref{fig3} are based on the assumption, common in the literature, that the CR densities go to zero at the boundaries of the CR confinement region, notably at an altitude of \zm\ above the disk. This could explain the differences in \zm\ with respect to studies of radio synchrotron emission, e.g., \cite{orlando2013}, if a sizable fraction of the emission is produced by interactions of CR electrons beyond the confinement region. 

Some important caveats apply to Fig.~\ref{fig3}. On one hand, the emissivities for HVCs and IVCs may be overestimated due to the presence of sizable amounts of ionized gas or undetected DNM. This however, would strengthen our conclusions on the decrease of the emissivity as a function of $z$. On the other hand the model predictions take into account only interactions between H and He nuclei in both CRs and the targets. A modest fraction ${\sim} 5\%$ of the emissivity is attributed to interactions with metals in the ISM \cite{mori2009}. While the relative abundance of He/H in the targets is not expected to vary because both elements are mostly primordial, a metallicity lower than the local value has been measured  in some HVCs \cite{wakker2001}. The related decrease in emissivity, however, is too small to explain the observed decrease as a function of $z$.

Our study demonstrates that \g-ray observations of HVCs and IVCs are a valuable resource to derive constraints on the CR densities in the halo of the Milky Way that are largely model independent. This can provide information on the structure and extent of the CR halo, as well as on the poorly understood problem of escape from the halo and how the outflow merges with CR densities in the local intergalactic space. It can also be compared to a larger class of models than those considered here, including models that entail CR-driven Galactic winds and anisotropic diffusion, non-uniform diffusion properties in the Milky Way, and more realistic representations of the geometry of the Galaxy. 

\section{Final remarks}

We have achieved the first detection of IVCs in \g-rays and set upper limits on other HVCs and IVCs with known locations in the halo of the Milky Way, thereby tracing for the first time the densities of the large-scale Galactic population of CR nuclei outside of the disk.

There is evidence at 97.5\% c.l. that the \g-ray emissivity per H~atom in the energy range from 300~MeV to 10~GeV, i.e., CR densities in the energy range  from ${\sim} 3$~GeV/nucleon to ${\sim} 200$~GeV/nucleon, decrease with distance from the Galactic plane. This provides new observational evidence of the Galactic origin of CRs in this energy range. 

The \g-ray observations of HVCs and IVCs can be compared to CR propagation models and provide unique information on CRs in the halo and their escape in the local intergalactic medium. 
The upper limit on the \g-ray emissivity in Complex~A at 22\% of the local value gives the most stringent constraint to date on the CR densities at a few kpc from the Galactic disk. The upper limit for the emissivity of the upper intermediate-velocity Arch hints at  a 50\% decline of the CR densities within 2~kpc from the disk. 

\acknowledgments

The \textit{Fermi} LAT Collaboration acknowledges generous ongoing support
from a number of agencies and institutes that have supported both the
development and the operation of the LAT as well as scientific data analysis.
These include the National Aeronautics and Space Administration and the
Department of Energy in the United States, the Commissariat \`a l'Energie Atomique
and the Centre National de la Recherche Scientifique / Institut National de Physique
Nucl\'eaire et de Physique des Particules in France, the Agenzia Spaziale Italiana
and the Istituto Nazionale di Fisica Nucleare in Italy, the Ministry of Education,
Culture, Sports, Science and Technology (MEXT), High Energy Accelerator Research
Organization (KEK) and Japan Aerospace Exploration Agency (JAXA) in Japan, and
the K.~A.~Wallenberg Foundation, the Swedish Research Council and the
Swedish National Space Board in Sweden.

Additional support for science analysis during the operations phase is gratefully acknowledged from the Istituto Nazionale di Astrofisica in Italy and the Centre National d'\'Etudes Spatiales in France.

This work was partially funded by NASA grant NNX13O87G.

\providecommand{\href}[2]{#2}\begingroup\raggedright\endgroup

\end{document}